\documentclass{appolb}
\usepackage{graphicx}
\usepackage{tikz}					
\usepackage{ifthen}					
\usepackage{tikz-3dplot}			
\usepackage{pgfplots}				
\usepackage{hyperref}               
\usepackage{amsmath}		       	
\usepackage{amsfonts}				
\usepackage{amssymb}				
\usepackage{mathtools}
\usepackage{cancel}					
\usepackage{dsfont}

\newcommand{\avg}[1]{\left\langle : #1:\right\rangle}
\newcommand{\W}{\mathcal{W}}

\renewcommand{\d}{\mathrm{d}}

\begin{document}
\title{Kinetic theory for massive spin-1 particles%
\thanks{Presented at Quark Matter 2022}%
}
\author{David Wagner, Nora Weickgenannt
\address{Institute for Theoretical Physics, Goethe University,\\ Max-von-Laue-Str.\ 1, D-60438 Frankfurt am Main}
\\[3mm]
{Enrico Speranza 
\address{Illinois Center for Advanced Studies of the Universe \& Department of Physics, University of Illinois at Urbana-Champaign,\\ Urbana, IL, 61801, USA}
}
}
\maketitle
\begin{abstract}
We calculate the Wigner function for charged spin-1 particles in inhomogeneous classical electromagnetic fields, going to first order in a power series in $\hbar$. The Boltzmann equation for the scalar distribution function obtained from this formalism agrees with previous calculations for spin-1/2 particles. In particular, we recover a Mathisson force of twice the magnitude, correctly reflecting the higher dipole moment of vector mesons. Evolution equations for vector and tensor degrees of freedom are obtained, and global equilibrium is discussed.
\end{abstract}
  
\section{Introduction}
Recently, an unexpectedly large tensor polarization of vector mesons has been observed in heavy-ion collisions \cite{ALICE:2019aid,Mohanty:2021vbt,STAR:2022fan}, which has not yet been satisfactorily explained, although there is already some theoretical progress \cite{Sheng:2022ffb, Weickgenannt:2022jes,Li:2022vmb, Wagner:2022gza}.
In this work, we employ methods similar to Ref. \cite{Weickgenannt:2019dks} to establish a quantum kinetic theory for massive spin-1 particles in classical electromagnetic fields up to first order in the Planck constant $\hbar$.

Our notation and conventions are: $A^{(\mu} B^{\nu)}:=A^\mu B^\nu +A^\nu B^\mu\;,A^{[\mu} B^{\nu]}\coloneqq A^\mu B^\nu -A^\nu B^\mu$, 
$E^{\mu\nu}\coloneqq k^\mu k^\nu /k^2$, $K^{\mu\nu}\coloneqq g^{\mu\nu}-E^{\mu\nu}$, $A\cdot B\coloneqq A^\mu g_{\mu\nu}B^\nu$, where $g^{\mu\nu}\coloneqq \text{diag}(1,-1,-1,-1)$.

\section{Field theory for massive vector particles}
\label{subsec:field_theory}
We consider the following Lagrangian for a complex Proca field $V^\mu$,
\begin{eqnarray}
 \mathcal{L}= \hbar \left(-\frac{1}{2}V^{\dagger\mu\nu} V_{\mu\nu} +\frac{m^2 }{\hbar^2} V^{\dagger\mu} V_\mu \right)-\frac{1}{4 }F^{\mu\nu} F_{\mu\nu} -   i q F_{\mu\nu} V^\mu V^{\dagger\nu}\;,
\end{eqnarray}
where $q$ is the electric charge, $F^{\mu\nu}\coloneqq \partial^{[\mu}A^{\nu]}$ is the electromagnetic field strength tensor, and $V^{\mu\nu}\coloneqq D^{[\mu} V^{\nu]}$, with $D^\mu\coloneqq \partial^\mu +(iq/\hbar) A^\mu$ being the covariant derivative in the fundamental representation. 
Note that the last term enhances the coupling of the Proca field to the electromagnetic field such that the gyromagnetic ratio is exactly equal to two \cite{Corben.1940, Ferrara:1992yc}.
The resulting equations of motion for the Proca field read
\begin{eqnarray}
D\cdot V&=& -i\frac{q \hbar}{m^2} J \cdot V\; ,\label{div_V}\\
\left(D\cdot D+\frac{m^2}{\hbar^2}\right)V^\nu &=&2\frac{iq}{\hbar}V_\mu F^{\mu\nu}-i\frac{q\hbar}{m^2}D^\nu \left(J\cdot V\right)\;,\label{EoM_V}
\end{eqnarray}
where $J^\mu \coloneqq  -iq \left[V^{\mu\nu}V^\dagger_{\nu} -V^{\dagger\mu\nu}V_{\nu} +\partial_\nu (V^{[\mu}V^{\dagger\nu]})  \right]$.

\section{Spin-1 quantum kinetic theory}
\subsection{Wigner function}
\label{sec:Wigner}
The Wigner function for charged vector particles in the presence of electromagnetic fields is defined as \cite{Vasak:1987um}
\begin{equation}
\label{def_W}
\W^{\mu\nu}(x,k)\coloneqq -\frac{2}{(2\pi\hbar)^4\hbar} \int \d^4 v e^{-ik\cdot v/\hbar} \avg{V^{\dagger\mu}_+ U_{+-}V^\nu_-}\;,
\end{equation}
where $V^\mu_{\pm}\coloneqq V^\mu\left(x_\pm\right)$, $x_\pm\coloneqq x\pm v/2$
and the gauge link $U_{+-}\coloneqq U(x_+,x_-)$ is defined as 
\begin{equation}
\label{gauge_link}
U(x,y)\coloneqq \hat{T}\exp\left\{ -\frac{iq}{\hbar} (x-y)\cdot \int_{-1/2}^{1/2} \d t A\left[\frac{x+y}{2}+t(x-y)\right]  \right\}\;,
\end{equation}
with $\hat{T}$ being the time-ordering operator.
Treating the electromagnetic field as classical, we may neglect the time ordering operator, as the fields commute at different times. 

Following Ref.~\cite{Vasak:1987um}, we introduce $\hat{K}^\mu \coloneqq  \hat{\Pi}^\mu +\frac{i\hbar }{2} \hat{\nabla}^\mu$, where $\hat{\Pi}^\mu\coloneqq k^\mu-\frac{\hbar q}{2} j_1(\Delta) F^{\mu\nu}\partial_{k,\nu}$ and $\hat{\nabla}^\mu \coloneqq  \partial^\mu-q j_0(\Delta) F^{\mu\nu}\partial_{k,\nu}$ denote the generalized momentum and derivative operators, respectively. Here $\Delta\coloneqq \frac{\hbar}{2}\partial_k\cdot \partial$ and 
$j_0(x)\coloneqq \sin(x)/x$, $j_1(x)\coloneqq [\sin(x)-x\cos(x)]/x^2$ are spherical Bessel functions. Note that the derivatives with respect to position contained in the $\Delta$'s act only on the field strength tensors. 

One can prove that the exact equations of motion for the Wigner function are given by
\begin{eqnarray}
\hat{K}_\nu \W^{\mu\nu}&=&\frac{q \hbar^2}{m^2}e^{-i\Delta}(\partial_\alpha F^{\alpha\nu}) \W^\mu_{\;\;\nu}\;, \label{cons_W}\\
\left(\hat{K}\cdot \hat{K}  -m^2\right)\W^{\mu\nu}&=&\left(\hat{K}_\alpha\hat{K}^\nu-i\hbar q e^{-i\Delta}F_\alpha^{\;\;\nu} \right)\W^{\mu\alpha}\label{eom_W}\;.
\end{eqnarray}
Equation \eqref{cons_W} follows from Eq.~\eqref{div_V} and reduces the number of independent degrees of freedom of the Wigner function, while Eq.~\eqref{eom_W} is a consequence of the field equation \eqref{EoM_V}.

\subsection{Decomposition and power counting}
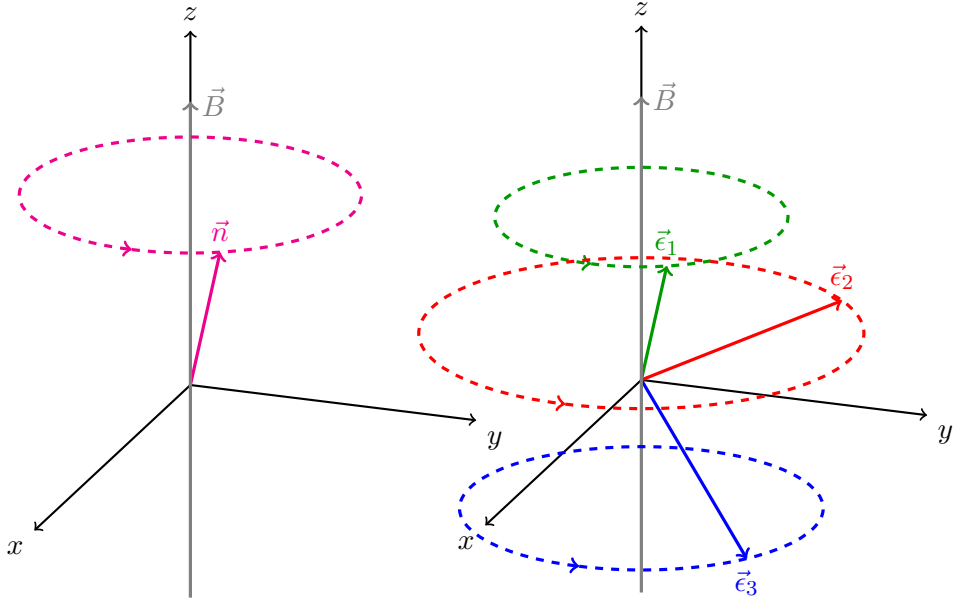
\begin{figure}[t]
\begin{minipage}{0.4\textwidth}
\centering
\tdplotsetmaincoords{70}{110}\begin{tikzpicture}[tdplot_main_coords]\draw[thick,->] (0,0,0) -- (6,0,0) node[anchor=north east]{$x$};\draw[thick,->] (0,0,0) -- (0,4,0) node[anchor=north west]{$y$};\draw[thick,->] (0,0,0) -- (0,0,5) node[anchor=south]{$z$};
\tdplotsetrotatedcoords{30}{40}{20}
\draw[very thick,color=magenta,tdplot_rotated_coords,->] (0,0,0) --(0,0,3.5) node[anchor=south]{$\vec{n}$};
\draw[very thick,white!50!black,->] (0,0,-3) --(0,0,4) node[anchor=west]{$\vec{B}$};
\tdplottransformrotmain{0}{0}{3.5}
\pgfmathsetmacro{\r}{\tdplotresx^2+\tdplotresy^2}
\tdplotdrawarc[very thick, dashed,color=magenta,->]{(0,0,\tdplotresz)}{sqrt{\r}}{0}{360}{anchor=north west,color=black}{}
\end{tikzpicture}
\end{minipage}
\begin{minipage}{0.4\textwidth}
\centering
\tdplotsetmaincoords{70}{110}\begin{tikzpicture}[tdplot_main_coords]\draw[thick,->] (0,0,0) -- (6,0,0) node[anchor=north east]{$x$};\draw[thick,->] (0,0,0) -- (0,4,0) node[anchor=north west]{$y$};\draw[thick,->] (0,0,0) -- (0,0,5) node[anchor=south]{$z$};
\tdplotsetrotatedcoords{30}{40}{20}
\draw[very thick,color=blue,tdplot_rotated_coords,->] (0,0,0) --(3,0,0) node[anchor=north]{$\vec{\epsilon}_3$};\draw[very thick,color=red,tdplot_rotated_coords,->] (0,0,0) --(0,3,0) node[anchor=south]{$\vec{\epsilon}_2$};\draw[very thick,color=green!60!black,tdplot_rotated_coords,->] (0,0,0) --(0,0,3) node[anchor=south]{$\vec{\epsilon}_1$};
\draw[very thick,white!50!black,->] (0,0,-3) --(0,0,4) node[anchor=west]{$\vec{B}$};
\tdplottransformrotmain{0}{0}{3}
\pgfmathsetmacro{\r}{\tdplotresx^2+\tdplotresy^2}
\tdplotdrawarc[very thick, dashed,color=green!60!black,->]{(0,0,\tdplotresz)}{sqrt{\r}}{0}{360}{anchor=north west,color=black}{}
\tdplottransformrotmain{0}{3}{0}
\pgfmathsetmacro{\r}{\tdplotresx*\tdplotresx+\tdplotresy*\tdplotresy}
\tdplotdrawarc[very thick, dashed,color=red,->]{(0,0,\tdplotresz)}{sqrt{\r}}{0}{360}{anchor=north west,color=black}{}
\tdplottransformrotmain{3}{0}{0}
\pgfmathsetmacro{\r}{\tdplotresx^2+\tdplotresy^2}
\tdplotdrawarc[very thick, dashed,color=blue,->]{(0,0,\tdplotresz)}{sqrt{\r}}{0}{360}{anchor=north west,color=black}{}
\end{tikzpicture}
\end{minipage}
\caption{Precession of the vector (left) and tensor (right) polarization around a magnetic field in the $z-$direction.}
\label{fig:BMT}
\end{figure}

We decompose the Wigner function as
\begin{equation}
\mathcal{W}^{\mu\nu}= \frac{g^{\mu\nu}}{4} \mathcal{V} -i\mathcal{S}^{\mu\nu}+\mathcal{T}^{\mu\nu}\;,
\end{equation}
where $\mathcal{V}\coloneqq \mathcal{W}^{\mu}_{\;\;\mu}$, $\mathcal{S}^{\mu\nu}\coloneqq i(\W^{\mu\nu}-\W^{\nu\mu})/2$ and $\mathcal{T}^{\mu\nu}\coloneqq g^{\mu\nu}_{\alpha\beta} \W^{\alpha\beta}$, where $g^{\mu\nu}_{\alpha\beta}\coloneqq g^{(\mu}_\alpha g^{\nu)}_\beta /2-g^{\mu\nu}g_{\alpha\beta}/4$. 
Solving Eqs.~\eqref{cons_W} and \eqref{eom_W} up to first order in $\hbar$, we obtain
\begin{eqnarray}
\mathcal{V}&=&\delta(k^2-m^2)V-\delta'(k^2-m^2)q\hbar F_{\alpha \beta} \Sigma^{\alpha\beta}\;,\\
\mathcal{S}^{\mu\nu}&=&\delta(k^2-m^2)\Sigma^{\mu\nu}-\delta'(k^2-m^2)q\hbar\left(\frac{1}{2} F^{\mu\nu} V - F_{\alpha}^{\;\;[\mu} t^{\nu]\alpha}\right) \;,\\
\mathcal{T}^{\mu\nu}&=&\delta(k^2-m^2)t^{\mu\nu}-\delta'(k^2-m^2)q\hbar g^{\mu\nu}_{\alpha\beta} F_{\gamma}^{\;\;\alpha} \Sigma^{\beta\gamma}\;.
\end{eqnarray}
The evolution of the on-shell parts $V, \Sigma^{\mu\nu}$ and $t^{\mu\nu}$  is determined by the following equations of motion,
\begin{eqnarray}
0&=&\delta(k^2-m^2)\bigg[ k\cdot \hat{\nabla}^{(0)} V+\frac{q\hbar}{2}(\partial^\gamma F^{\alpha\beta})\partial_{k,\gamma}\Sigma_{\alpha\beta}  \bigg]\label{Boltzmann_F01}\;,\\
0&=&\delta(k^2-m^2)\bigg[k\cdot\hat{\nabla}^{(0)} \Sigma^{\mu\nu}- qF_\rho^{\;\;[\mu} \Sigma^{\nu]\rho}\nonumber\\
&&-\frac{q\hbar}{2}(\partial^\gamma F_\alpha^{\;\;[\mu})\partial_{k,\gamma}\left(t^{\nu]\alpha}+\frac{1}{4}g^{\nu]\alpha} V\right)-\frac{q\hbar}{2m^2}J_\alpha \left(t^{\alpha[\mu}+\frac14 g^{\alpha[\mu} V\right)k^{\nu]} \bigg]\label{Boltzmann_S01}\;,\qquad\;\\
0&=&\delta(k^2-m^2)\bigg[ k\cdot \hat{\nabla}^{(0)} t^{\mu\nu} +qF_\alpha^{\;\;(\mu}t^{\nu)\alpha}\nonumber\\
&& +\frac{q\hbar}{2}g^{\mu\nu}_{\alpha\beta} (\partial^\gamma F^\alpha_{\;\;\rho})\partial_{k,\gamma} \Sigma^{\beta\rho} -\frac{q\hbar}{2m^2}J_\alpha \Sigma^{\alpha(\mu}k^{\nu)}\bigg]\label{Boltzmann_Fmunu01}\;.
\end{eqnarray}
Here, we defined $\hat{\nabla}^{(0)}_\mu \coloneqq \partial_\mu -q F_{\mu\nu} \partial_{k}^{\nu}$ as the expansion of the operator $\hat{\nabla}_\mu$ to first order in $\hbar$.
These on-shell Boltzmann-Vlasov-like equations closely resemble those derived in Ref.~\cite{Weickgenannt:2019dks} for spin-1/2 particles, however, with a key difference. Namely, the parts of the quantity $\mathcal{T}^{\mu\nu}$ that are orthogonal to the four-momentum $k^\mu$ are related to the tensor polarization of the particles \cite{Leader:2001, Wagner:2022gza}, which is a genuine spin-1 property that is absent for spin-1/2 particles. Thus, when setting $\mathcal{T}^{\mu\nu}=0$, we may compare the resulting expressions to Eq.~(60) in Ref.~\cite{Weickgenannt:2019dks}. It becomes clear that, as expected, Eq.~\eqref{Boltzmann_F01} differs by a factor of two in the term resembling the Mathisson force, while Eq.~\eqref{Boltzmann_S01} features a factor of one half in the force term. This is consistent with the fact that the spin magnitude of vector mesons is twice as large as that of spin-1/2 particles.
Furthermore, it should be noted that the components of the Wigner function parallel to the momentum, i.e., the seven independent quantities $k_\mu \mathcal{S}^{\mu\nu}$ and $k_\mu \mathcal{T}^{\mu\nu}$, are fixed by Eq.~\eqref{cons_W}. In particular, to zeroth order in $\hbar$, we have $k_\mu \mathcal{S}^{\mu\nu}=k_\mu \mathcal{T}^{\mu\nu}=0$.
At this order, Eq.~\eqref{Boltzmann_S01} is equivalent to the Bargmann-Michel-Telegdi (BMT) equation \cite{Bargmann:1959gz} for the classical ``spin vector'' $n^{\mu}\coloneqq -(1/2)\epsilon^{\mu\nu\alpha\beta}(k_\nu/m)\Sigma_{\alpha\beta}$, while Eq.~\eqref{Boltzmann_Fmunu01} gives the BMT equations for the three (spacelike) eigenvectors of $t^{\mu\nu}$, which we denote by $\epsilon_i^\mu$, where $i\in\{1,2,3\}$ and $\epsilon_i\cdot k=0$. These equations describe the precession around the magnetic field in the particle rest frame, cf. Fig.~\ref{fig:BMT}.

\section{Global equilibrium}
Assuming the standard form of the collision term, the equilibrium distribution function can be parametrized as $f^{e}\coloneqq \left[ \exp({g^{e}})-1\right]^{-1}$, where $g^e$ denotes a combination of the conserved quantities (charge and four-momentum), i.e., $g^e \coloneqq \beta\cdot k-e\alpha$.
Here, $e\in\{+,-\}$ distinguishes particles and antiparticles. The quantities $\alpha$ and $\beta^\mu$ are the Lagrange multipliers associated to the conserved charges. Note that we assume that polarization effects arise at first order in $\hbar$. This implies that, compared to the discussion in Ref. \cite{Weickgenannt:2019dks}, we do not have to add a term $\sim \Sigma^{\mu\nu}$ corresponding to the conservation of total angular momentum.
From the Boltzmann equation \eqref{Boltzmann_F01} it follows that in global equilibrium the following conditions have to hold \cite{Weickgenannt:2019dks},
\begin{equation}
    \partial^\mu \alpha= qF^{\mu\nu}\beta_\nu \;, \quad \partial^{(\mu}\beta^{\nu)}=0\;.
\end{equation}
Employing Eqs.~\eqref{Boltzmann_F01}--\eqref{Boltzmann_Fmunu01}, the components of the Wigner function in global equilibrium to order $\mathcal{O}(\hbar)$ are given by
\begin{eqnarray}
\mathcal{T}^{\mu\nu}+\frac{g^{\mu\nu}}{4}\mathcal{V} &=& \delta(k^2-m^2)\left(K^{\mu\nu}V+\hbar\Phi^{\mu\nu}\right)\;,\\
\mathcal{S}^{\mu\nu}&=&\hbar\delta(k^2-m^2)\left[ -\left(\varpi^{\mu\nu}+\frac12 E_\alpha^{\;\;[\mu}\varpi^{\nu]\alpha}\right)V' \right.\nonumber\\
&&\left.+\frac{q}{2m^2}\left( F^{\mu\nu} +2 E_\alpha^{\;\;[\mu} F^{\nu]\alpha}\right)V   + \Xi^{\mu\nu}\right]\nonumber\\
&&+\hbar\delta'(k^2-m^2) qF_\alpha^{\;\;[\mu}K^{\nu]\alpha} V\;,
\end{eqnarray}
where 
\begin{equation}
V\coloneqq  \frac{1}{(2\pi\hbar)^3} \frac13 \sum_{e} \Theta(ek^0) f^{e}\;,\quad V'\coloneqq  \frac{1}{(2\pi\hbar)^3} \frac13 \sum_{e} \Theta(ek^0) \frac{\partial f^{e}}{\partial g^{e}}\;,
\end{equation}
and we defined the thermal vorticity $\varpi^{\mu\nu}\coloneqq -(1/2)\partial^{[\mu} \beta^{\nu]}$.
The terms $\Phi^{\mu\nu}$ and $\Xi^{\mu\nu}$ follow the BMT equations and are otherwise unconstrained, i.e., they provide no information about the medium.
On the other hand, the contribution $\sim \varpi^{\mu\nu}$ encodes the massive analogue of the axial chiral vortical effect, while the terms $\sim F^{\mu\nu}$ determine the chiral separation effect \cite{Kharzeev:2015znc,Avkhadiev:2017fxj,Yamamoto:2017uul,Huang:2020kik}. This can be seen when considering the axial current
\begin{equation}
    J_A^\mu = \frac{1}{2m}\int \d^4 k\, \epsilon^{\mu\nu\alpha\beta} k_\nu \mathcal{S}_{\alpha\beta}\;.
\end{equation}
Spin-1 particles, in addition to their magnetic dipole moment, possess an electric quadrupole moment which the electromagnetic field can couple to. This may induce a tensor polarization even in global equilibrium, but this effect is expected to occur at second order in $\hbar$ \cite{Delgado-Acosta:2012dxv}.
In contrast, away from equilibrium, a nonvanishing tensor polarization is induced by the shear-stress tensor of the medium \cite{Wagner:2022gza}. 
\newline\newline
{\it Acknowledgements} ---
The authors thank D. H. Rischke for enlightening discussions. The work of D.W. and N.W.\ is supported by the
Deutsche Forschungsgemeinschaft (DFG, German Research Foundation)
through the Collaborative Research Center CRC-TR 211 ``Strong-interaction matter
under extreme conditions'' -- project number 315477589 - TRR 211 and by the State of Hesse within the Research
Cluster ELEMENTS (Project ID 500/10.006). D.W. acknowledges support by the Studienstiftung des deutschen
Volkes (German Academic Scholarship Foundation).

\bibliographystyle{h-physrev}
\bibliography{biblio_paper_long}

\end{document}